# SLA-Awareness for AI-assisted coding

Kishanthan Thangarajah*, Arthur Leung*, Boyuan Chen*, Ahmed E. Hassan°
cse@huawei.com
* Centre for Software Excellence, Huawei Canada ° Queen's University, Canada

**ABSTRACT**

The integration of AI-assisted coding tools within development environments drastically reduces development time, and allows developers to focus more on creative and critical aspects of software engineering through the use of Code Large Language Models (CodeLLMs) [52]. These coding assistants automate repetitive and time-consuming coding tasks such as code generation, code completion, code summarization, and code translation. Responsiveness is a crucial requirement of these coding assistants to maintain real-time interactivity, such that their use does not impede the developers' workflows. Different coding tasks have unique input/output characteristics and latency requirements: Time-To-First-Token (TTFT) latency is essential for code completion tasks, while End-To-End (E2E) latency is crucial for code translation tasks. Managing these varying requirements simultaneously while optimizing resource usage poses significant challenges. Existing work adopts the Model-as-a-Service paradigm for serving individual CodeLLMs, but cannot effectively manage latency requirements of concurrent coding tasks and sequences of CodeLLM inference calls, due to a lack of end-to-end latency awareness. Another challenge after meeting latency requirements is keeping resource utilization high, when the serving system is deployed on a shared cluster environment. To address these challenges, we propose **Coding Assistant Task Orchestrator (CATO)**, a runtime system designed to serve a diverse assortment of coding tasks while meeting latency requirements and maximizing resource utilization. Our experiments demonstrate that when all types of coding tasks were served simultaneously, for TTFT-critical tasks (code completion, code generation), CATO improves overall Goodput rate and resource utilization by up to **10%** and **41.1%**, respectively. P95 E2E latency was also reduced by **18%** for code summarization tasks, and P95 TTFT for code generation tasks were reduced by **14%** compared against state-of-the-art systems.

## 1 INTRODUCTION

CodeLLMs are a special type of LLMs which are specifically trained on extensive amounts of code artifacts and documentation to handle various coding tasks [52]. AI-assisted coding through the use of CodeLLMs is revolutionizing software development by enhancing both productivity and innovation [11, 13, 15, 20, 31]. Key capabilities of CodeLLMs include generating source code (code generation), providing real-time suggestions (code completion), and accelerate repetitive tasks (code refactoring, code translation, and code summarization). Development time and debugging overhead can be effectively reduced in these use cases, allowing developers to concentrate more on the creative and strategic aspects of their projects, and overall makes programming more accessible to individuals in all levels of expertise.

Although many CodeLLMs are proprietary and externally hosted (Codex by OpenAI [37], Claude by Anthropic [3]), enterprises still wish to use in-house trained/fine-tuned CodeLLMs for three primary reasons: Data privacy, data customizability, and performance guarantees. Firstly, companies must abide by privacy regulations such as GDPR when operating LLMs in certain regions [24, 38]). Secondly, enterprises may wish to customize the capability of their CodeLLMs, by serving fine-tuned versions of them, and Mixture of Experts (MoE) models for tailored use cases. For instance, a fine-tuned version of CodeLlama [44] is used internally at Meta [16] specifically for its CodeCompose tool. Lastly, in-house hosting offers performance guarantees, which would not be possible by outsourcing: Cursor team has used a MoE model to handle large source code files and long input context length while reducing inference latency [18].

On top of using custom-hosted models, users typically have the option to choose a CodeLLM to use with coding assistant tools (Copilot [10], Cursor [12]). The same coding assistant can also be invoked for different coding tasks across many users simultaneously. For example, one developer using Cursor AI editor with CodeLlama might call for code generation, while another developer is performing code refactoring using the same model. Meanwhile, other developers could be utilizing the same model for real-time code completion. Even though developers may be unaware of how the coding assistant manages these concurrent tasks to the same CodeLLM, they still expect quick and efficient responses to maintain their own development workflow.

Keeping latencies low is a critical Service Level Agreement (SLA) requirement for coding assistants to work well in real-time integrated development environments (IDEs), as developers need quick feedback for their tasks. User complaints about delays when using coding assistants are prevalent [2, 6, 7, 14, 29, 33, 40]. This also applies for paid subscription users [9]). These issues come down to how efficiently the underlying infrastructure for coding assistants can schedule and scale, to handle a concurrent assortment of coding tasks.

At a high level, this involves provisioning CodeLLM instances, coordinating multiple CodeLLM calls for generating, verifying, and refining outputs, as well as handling additional processes like fetching documents or examples from databases in the coding tasks which involve them, such as Retrieval Augmented Generation (RAG).

Currently, CodeLLMs are served using *Model-as-a-Service* systems, however they struggle to deliver high Goodput and high resource utilization. This is mainly because they do not (a) evaluate SLA compliance based on overall performance of the entire coding task, (b) manage simultaneous coding tasks with conflicting SLA demands, such as low TTFT for code generation and high throughput for code translation, and (c) balance Goodput and hardware utilization, which require effective autoscaling and to have resources at the ready under varying workloads.

We introduce **CATO**, an SLA-Aware runtime which can efficiently orchestrate CodeLLM-based coding tasks with SLA awareness. With CATO, we propose two novel SLA-Aware algorithms for both scheduling and scaling to serve CodeLLM requests. We experimentally evaluated this system by load testing on an assortment of coding tasks, based on their characteristics and SLA requirements. Our experiments show that, compared to existing state-of-the-art baseline (Ray Serve [43]), our system achieves **10%** higher Goodput, **41.1%** higher utilization, up to **18%** reduction in P95 E2E latency in code summarization tasks, and **14%** reduction in P95 TTFT latency in code generation tasks. The rest of the paper is organized as follows. In Section 2, we give the background on AI-assisted coding tasks and their characteristics. In Section 3, we present challenges faced when serving different coding tasks and we proposed our solution in Section 4. In Section 5, we show our evaluation methods and accompanying discussion. In Sections 6, 7, 8, we discuss threats to validity, related work and conclude the paper.

## 2 BACKGROUND

Latency is a crucial SLA aspect for coding assistants that have been used in IDEs, where tasks have stringent responsiveness requirements. Developers expect immediate responses in particular coding tasks. For instance, in a code completion task, developers cannot afford to wait several seconds for a suggestion while typing; it must keep pace with the developer's typing speed. This requirement is prevalent in works such as CodeCompose [16], where developers expected the assistant to return code suggestions within milliseconds, including multi-line suggestions. Similarly, to reduce the latency in showing vulnerable code as the developer is typing, DeepVulGuard [45] used a small transformer-based model with a fewer number of parameters. The Cursor team [18] used a MoE (Mixture of Experts) model to reduce the overall latency when processing large input contexts and source code files. These examples show the extensive efforts into reducing latency to improve responsiveness for coding assistants in IDEs.

Different coding tasks will have different input and output characteristics as depicted in Figure 1. The specific nodes within each task are also decomposed and shown on Figure 2, consisting of CodeLLM calls (in blue) and non-CodeLLM calls. These four coding tasks are inspired from Open Platform for Enterprise AI (OPEA) Coding Examples [35, 36] and used as reference implementations in our experiments (Section 5).

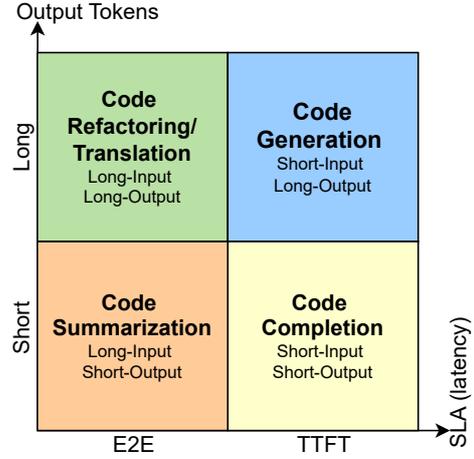

**Figure 1: Latency and output characteristics of coding tasks**

Likewise, users' expectations vary for different coding tasks. TTFT is crucial for tasks such as code completion [16] and code generation, while E2E latency is critical for tasks such as code refactoring/translation and code summarization. We describe these 4 tasks with their input/output characteristics and their latency requirements as follows:

- **code generation (Short-Input Long-Output):** Code generation is a task where the developer interactively provides specifications to the CodeLLM to generate code artifacts such as entire functions, classes, or even full scripts [5, 35]. Typically, in code generation tasks (Figure 2a) the first stage will fetch some relevant examples or documents from the database, then the generation prompt will be augmented with this additional context before sending to CodeLLM [21, 23, 47]. Optionally, to reduce compilation errors or improve output, the generated output can be validated for syntax and logical errors [23, 39] or improved with feedback [30] with another CodeLLM call and supporting code checker tools. The TTFT latency should be minimal to allow developers to explore and iterate rapidly on the generated code[5], whereas end-to-end latency is not as crucial so long as the token generation speed achieves at least the average human reading speed.
- **code completion (Short-Input Short-Output):** Code completion is a subset of code generation but involves providing real-time edit suggestions as developer types the code on the same line, mainly for acceleration of the coding process. In this task (Figure 2b), the partial code snippet and the context (relevant examples and documents) are first fetched and then sent to a CodeLLM to generate possible suggestions [8, 16, 49]. Real-time code completion requires ultra-low latency, with response times in the millisecond-range. This is required to keep pace with the developer's typing speed, providing seamless, and instant suggestions.

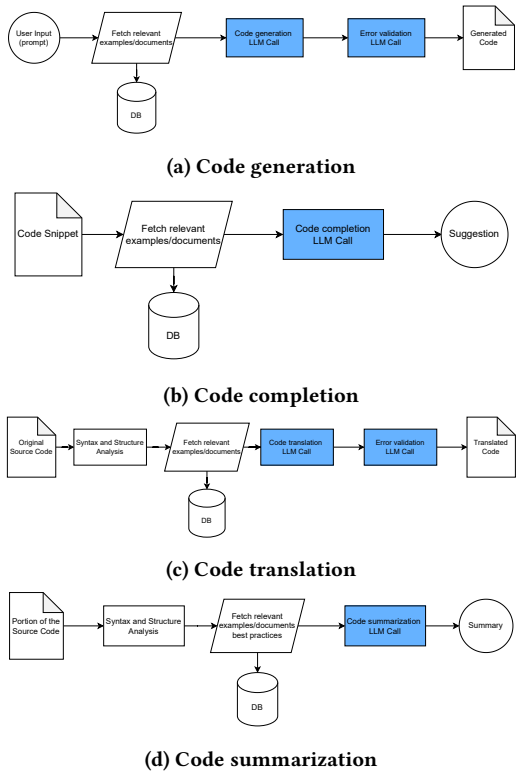

Figure 2: Different coding tasks (blue boxes represent CodeLLM calls)

- **code translation (Long-Input Long-Output):** Code translation involves converting code artifacts from one programming language to another [36]. This operation can involve some code snippets or entire repositories, helping developers to adapt their code to different environments and leverage the strengths of various programming languages. In this task (Figure 2c), the original source code to be translated will first be analyzed to understand the structure and logic [51]. Using this information, relevant examples could be fetched from a database which will then be sent to the CodeLLM to translate. To reduce compilation errors, the translated output from the LLM can be validated for syntax and logical errors using compiler tools or another CodeLLM call [50]. For code translation tasks, users typically would wait until the whole translated output becomes available. The output may additionally be accompanied by human readable comments and explanations.

- **code summarization (Long-Input Short-Output):** Code summarization involves creating concise and meaningful descriptions of code artifacts. This task helps developers understand and document previously unseen code quickly, improving code readability and maintainability. In this task (Figure 2d) users will select a portion of the code (code snippets, functions, or entire modules) and send these artifacts to the model as input. During the first stage, the code to be summarized will be analyzed to understand the structure and logic. This information will be used to fetch relevant documents and examples to augment the summarization call, then the next stage will call a CodeLLM with the instruction to generate a summary of the given code [1].

Production grade frameworks, such as Ray Serve [43], Fast Chat [17] and Triton [34], are used to deploy these models. Since a coding task can contain multiple CodeLLM and non-CodeLLM calls (code translation from Figure 2c), there is a need to decompose each coding task and invoke each of the stages individually. This requires an additional layer to manage the end-to-end execution of each coding tasks. However, there are no existing work which can orchestrate this execution to ensure SLA compliance and resource utilization. Additionally, the heterogeneity of the requests in different coding tasks, in terms of input lengths, output lengths, and expected SLA (TTFT and E2E latencies) present challenges to the coding assistant's deployment infrastructure, potentially impacting their responsiveness and accuracy.

Ensuring that coding assistants meet these diverse expectations requires sophisticated optimization techniques and a deep understanding of the specific requirements of each task. Developers rely on these coding assistants not only for their functionality but also for their efficiency and speed, which directly affect their workflow and productivity. Therefore, continuous improvements in latency, processing speed, and suggestion accuracy are essential for maintaining a positive user experience and maximizing the benefits of AI-assisted coding in software development.

## 3 CHALLENGES

Ensuring optimal performance and reducing latency across all coding assistant tasks involves several considerations. First, it's essential to manage the end-to-end performance of each coding task, which includes coordinating multiple CodeLLM calls for generating, verifying, and refining outputs. This also requires efficient provisioning of CodeLLM instances, and handling additional processes such as fetching documents or examples from databases. Maintaining computational resources and ensuring low latency is crucial to providing real-time performance in IDEs. Failure to ensure could result in bad development experiences and low user retention [2, 6, 7, 14, 29, 33, 40]. We have outlined three key challenges below, which coding assistants must overcome to effectively serve coding tasks.

**Model-as-a-Service serving cannot capture overall perceived performance of coding tasks:** One CodeLLM may be used in multiple coding tasks, each with diverse SLA requirements. For instance, in Cursor and Copilot, users have the ability to choose the base CodeLLM [10, 12] for the coding assistant to perform an assortment of tasks during development time. Consider an example serving one code generation and one code completion task: the first demands high throughput and the second requires low latency in their LLM invocations, but a model replica cannot be served with both optimizations simultaneously enabled, without knowing ahead of time this workflow information, as it only receives the prompt at the service level. Therefore, SLA compliance should be viewed from the perspective of the entire coding task (end-to-end), rather than serving one individual CodeLLM replica. Failure to mitigate this challenge can result in high latency variations when serving different coding tasks simultaneously, leading to low user retention. A study from Meta shows developers have opted to avoid using CodeCompose when it had an unacceptable latency delay [16]. To overcome this, CodeCompose developers had to apply techniques to the model hosting infrastructure such as batching, streaming, and queuing priority to reduce latencies.

**Conflicting SLA demands when serving an assortment of coding tasks:** It's necessary to manage simultaneous requests for different coding tasks and ensure that they do not interfere with each other. Code generation and code completion requires TTFT latency to be low, while other coding tasks (such as code translation and code summarization) requires higher throughput and take longer time to finish. Failure to address this can lead to inconsistent response times in real-time tasks and inefficient handling of background tasks. Furthermore, there are many reasons to provision model replicas across different geographic availability zones (GDPR [24, 38]) which affects latency. Provisioning a CodeLLM is an expensive operation compared to traditional applications, because there may be numerous versions of fine-tuned LLMs to be served for A/B testing [41]. Failure to mitigate this challenge results in poor responsiveness of the code assistant. For example, an AWS CodeWhisperer blog post mentions that its service is hosted in us-east-1, so users of this code assistant that are far from this availability zone may experience delays [32].

**Poor Goodput in favor of high utilization of hardware:** With latency-based SLAs, a common metric to understand a serving system's performance is *Goodput*, which is defined as the fraction of workflow requests meeting the SLA target. The granularity of control needed to handle uncertainties due to sudden changes (spiky workload, heterogeneous input and output tokens from inference requests) in CodeLLMs for different coding tasks also influences utilization. For instance, Cursor and Copilot users have experienced long delays due to queuing and high load [6, 7] as the CodeLLM service's replicas are over-utilized. If the arrival rate of requests exceeds the service rate across all worker replicas, then naturally the queue length grows as stipulated by Little's law [27], unless more replicas can be provisioned. Conversely, requests may be serviced instantly by having many replicas provisioned at idle with empty queues, but from a utilization/ efficiency standpoint for cluster operators, this is a wasteful practice. Therefore, many serving systems use autoscaling to provide elasticity and gauge compute demand, which can provision just enough replicas to strike a balance between Goodput and Utilization. Even with autoscaling, it is difficult for cluster operators to decide whether separate coding task deployments should use a dedicated cluster or share one same cluster utilizing the underlying hardware resources. On one hand, low resource utilization would still be a concern if separate clusters are formed for each of these deployments, as some tasks are seldom invoked. On the other hand, when tasks involving common CodeLLMs share the same underlying cluster, they will compete for hardware resources, as each of the services has its own scaling/scheduling configurations.

To address these challenges, a self-adapting SLA-aware execution runtime is needed. The runtime should be capable of both request workload routing (scheduling) and resource provisioning (scaling) to effectively meet the SLA goals of different coding tasks while serving them simultaneously for many users.

## 4 SLA-AWARE CODING ASSISTANT TASK ORCHESTRATOR

The challenges listed in Section 3 necessitates a reference architecture for a LLM serving system, capable of meeting these coding tasks' specific performance requirements. We introduce the CATO architecture in Section 4.1, with relevant metrics and design choices. The rest of the section is organized as follows: Section Section 4.2 describes in detail the responsibility of each core component of CATO. Section 4.3 describes the SLA-aware scheduling algorithm in which CodeLLM requests are scheduled onto LLM replicas of the cluster. Section 4.4 describes the SLA-aware scaling algorithm in which LLM replicas are provisioned in the cluster to handle varying loads.

### 4.1 CATO Architecture Overview

The high level architectural requirement of this system is to execute any type of multi-step coding task, which can be conceptualized as workflows, with the goal of meeting latency SLAs in all tasks. It should scale to support many concurrent requests, and monitor the following metrics: the top portion of Figure 3 describes the latency metrics of interest to be observed each time a coding task is invoked by a user, namely TTFT, E2E, and Slack time. Slack time is the time allotted for each node to execute, such that the sum of all slack time on the critical path does not exceed SLA. This Slack time may be fixed or determined arbitrarily, which will be elaborated when the *Profiler* component is introduced.

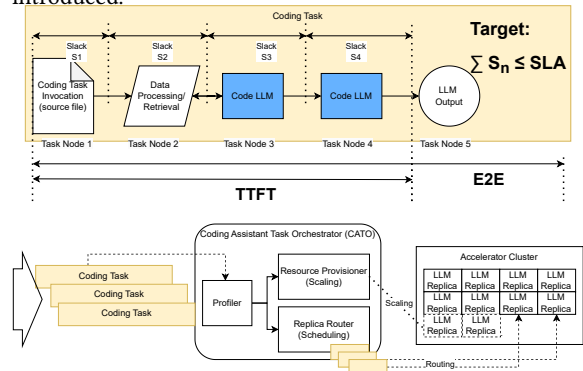

Figure 3: Top: task-level latency measurements within a coding task. Bottom: CATO system components processing incoming coding task requests

The bottom half of Figure 3 describes the system architecture with CATO as the control plane performing actions such as scheduling requests onto each replica and provisioning additional replicas on the cluster. CATO itself is further comprised of 3 core components: the *Profiler*, *Resource Provisioner*, and *Replica Router*.

### 4.2 Core components in CATO

**Profiler**: This component is responsible for measuring baseline latencies for each coding task. As each task has a different sequence of task nodes and characteristics, establishing this baseline is crucial to understand whether a given SLA requirement is realistic in the first place, in relation to execution on real hardware on the cluster at runtime. Based on this collected profile, an estimate of the allotted slack value for each task node is stored for the task. For example, if a coding task's nodes A,B,C take 5s, 15s, 5s to execute E2E, and the user provided SLA was 50s, then the Profiler may determine allotted slack proportionally to be 10s, 30s, 10s.

**Replica Router**: This component examines incoming task requests and decomposes them into individual task nodes. Some task nodes may be for database retrievals or data (pre-)processing, others may be CodeLLM invocations. **Scheduling decisions are performed here** to route these CodeLLM invocations to the best available LLM replica in the cluster, by continually monitoring the state of each replica. For instance, the state of each replica may either be *busy* or *idle*, where the former means the LLM is actively performing inference for a CodeLLM invocation. By employing different scheduling policies in this component, various metrics and heuristics may be monitored and used to drive scheduling decisions. As an example, Ray Serve selects 2 replicas randomly and routes the request to the replica with shortest queue, referred to as *power-of-2 scheduling* [43]. Other load balancing methods such as round-robin are also commonly employed.

**Resource Provisioner**: This component continually monitors the cluster's available accelerator hardware resources. **Scaling decisions are performed here** for creating and destroying replicas. Each machine in the cluster may only host a few accelerators (i.e. up to 8 cards per machine), so certain scaling decisions must follow allocation constraints. For instance, to load a 34B model, it requires the memory capacity of 2 accelerator cards on the *same machine*, and not any 2 accelerator cards across the cluster belonging to different machines. This component handles such constraints when acquiring/ releasing resources for creating/destroying LLM replicas. In popular distributed systems with autoscaling capabilities such as Ray Serve[43] and Kubernetes[22], request queue length or CPU utilization metrics are used as heuristic signals to indicate whether the system needs to scale up to meet incoming request demand.

### 4.3 SLA-Aware Scheduling Algorithm

Algorithm 1 describes a SLA-Aware method to schedule a CodeLLM request to a LLM replica, accompanied with the following numbered explanations. It first measures SLA at the task level ①. This is unique compared to existing state-of-the-art systems as they do not sufficiently capture the SLA requirements of coding tasks when scheduling. Primitive heuristics such as load, memory usage or queuing length cannot capture the dynamic nature of CodeLLM execution times, at each stage/ node of the task, we account for this

**Algorithm 1** SLA-Aware Scheduling

1: **procedure** SCHEDULE(invocation, model_id, replica_manager)
2:  min_remaining_time ← INT.MAX
3:  task_sla ← invocation.task.target_latency ①
4:  time_already_spent ← current_time - task_start_time
5:  slack ← task_sla - time_already_spent ②
6:  remaining_completion ← get_remaining_expected_completion()
7:  is_replica_selected ← False
8:  available_replicas ← replica_manager.get_available_replicas(model_id)
9:  **for** replica **in** available_replicas **do**
10:   wait_time ← replica.get_wait_time() ③
11:   **if** slack < 0 **and** wait_time < min_remaining_time **then** ④
12:    min_remaining_time ← wait_time
13:    chosen_replica ← replica
14:    is_replica_selected ← True
15:    priority ← 0
16:   **end if**
17:   expected_time ← wait_time + remaining_completion
18:   **if** expected_time < slack **and** wait_time < min_remaining_time **then** ⑤
19:    min_remaining_time ← wait_time
20:    chosen_replica ← replica
21:    is_replica_selected ← True
22:    priority ← 2
23:   **end if**
24:   **if** not replica_selected **and** wait_time < min_remaining_time **then** ⑥
25:    min_remaining_time ← wait_time
26:    chosen_replica ← replica
27:    is_replica_selected ← True
28:    priority ← 1
29:   **end if**
30:  **end for**
31:  chosen_replica.reorder_invocations(invocation, priority)
32:  **return** (chosen_replica, priority)
33: **end procedure**

using "slack" which is the remaining allowed time for the current invocation ②. The expected queuing time on each replica is calculated by iterating through the elements remaining in the queue and calculating each of their expected time to complete ③. This is based on values provided by the *Profiler* as well as execution times of similar coding tasks. If the slack is already violated (i.e., the request has already exceeded its allowed time), it selects a replica which has the shortest remaining queue time, and assigns the execution with the highest priority of 0 ④. If sufficient slack is available, select the replica which can both meet this slack and has the shortest remaining queue time, and assign it a lower priority of 2 ⑤. If there are no replicas which can meet slack, select replica has the shortest remaining queue time, and assign it a higher priority of 1 ⑥.

In summary, the Replica Router selects a replica based on "slack time", where the replica with shortest queuing delay AND meeting slack will have the request scheduled onto it. Requests with violated slack or nearing violation are reordered within the priority queue, when no other replicas are available.

**Algorithm 2** SLA-aware Resource Provisioning

```
1:  procedure SCALE(invocation, model_id, replica_manager)
2:      if model_id not in replica_manager.model_replicas then
3:          initialize first replica for model_id
4:          return
5:      end if
6:      if number of replicas for model_id ≥ max_replicas then
7:          // Max replicas reached, do nothing
8:          return
9:      end if
10:     node_target_latency ← get_execution_latency() + cpu_gpu_loading_time
11:     last_elapsed_metric ← get_last_metric_for_node()
12:     task_sla ← invocation.task.target_latency  ①
13:     time_already_spent ← current_time - task_start_time
14:     remaining_slack ← task_sla - time_already_spent  ②
15:     remaining_completion ← get_remaining_expected_completion()
16:     idle_replicas ← get_idle_replica_ids()
17:     if remaining_slack <0 and idle_replicas exist then
18:         return
19:     end if
20:     if remaining_completion > remaining_slack then
21:         exceeded_by ← last_elapsed_metric - node_target_latency
22:         if (exceeded_by/node_target_latency) ≥ max_exceeded_proportion then  ③
23:             exceeded_times ← exceeded_times + 1  ④
24:         end if
25:     else
26:         return
27:     end if
28:     num_existing ← count of existing replicas
29:     if (exceeded_times > max_exceeded_times) and (exceeded_times > count of idle_replicas) then  ⑤
30:         required ← exceeded_times - count of idle_replicas
31:         delta ← min(max_replicas - num_existing, required)
32:         scale_to_desired_replicas(num_existing + delta)
33:         exceeded_times ← 0
34:     else
35:         // Do nothing
36:     end if
37: end procedure
```

### 4.4 SLA-Aware Scaling Algorithm

The autoscaling policy described in Algorithm 2 considers SLA by apportioning each node with some slack time such that the total slack of the task's critical path would not exceed SLA. When the coding task is invoked at runtime, this slack is measured by subtracting one node's execution time ② from its allotted time obtained during profiling ①. A negative slack indicates that the execution took longer than expected for that node, which constitutes a slack violation ③. A positive slack means the execution completed ahead of the expected time, and therefore the request can afford some queuing delay before the next node in the task executes, thus the scheduler can have take some liberties in reassigning its priority in the replica's queue. All slack violations in a time period across all nodes are counted ④. When a threshold is exceeded, the **Resource Provisioner** creates an additional number of replicas of the respective CodeLLM proportional to slack violation count, to minimize future slack violations

⑤. Idle replicas are destroyed after a predefined timeout but omitted for brevity.

## 5 EVALUATION

In this section, we evaluate CATO against two baselines and present our findings. We addressed four Research Questions (RQs) to demonstrate the performance of all three systems when serving different types of coding tasks simultaneously. Below are the descriptions of the scheduling and scaling policies of CATO and the baselines used in the experiments.

- **CATO:** uses SLA-Aware scaling and SLA-Aware scheduling.
- **Ray Serve:** uses queue length based scaling and power of two choices based scheduling
- **Round Robin:** added to compare the scheduling experiments mainly. For scaling, it uses maximum queue length policy (same as Ray Serve scaling) and round robin based scheduling (used commonly in production grade LLM serving systems like Triton [34].)

### 5.1 Experiment Setup

Two Atlas 800 model 9000 machines form the cluster used in all experiments: one with 8 x Ascend 910B4 and the other with 4; a total of 12 accelerator cards. Each accelerator has 32GB memory to load models and perform inference. 10Gbps networking connected both machines, and the model weights storage was hosted using a mounted NFSv4 share. For all three systems, vLLM-0.4.2 was used as the underlying inference engine.

### 5.2 Metrics

Across all experiments, we evaluate on three key performance metrics.

**Goodput:** *Goodput* is defined as the number of requests meeting the SLA divided by the total number of requests. In the Goodput measurement figures in each RQ, each color represents different scheduling policies under test (CATO, Ray Serve, and Round Robin).

**Latency:** For code completion and code generation, we measured TTFT latency, defined as the difference between the timestamp when the first token was generated and the timestamp when the task was invoked. For code summarization and code translation, we measured the E2E latency, defined as the difference between the timestamp when the last token was generated and the timestamp when the task was invoked. P95 and P99 refer to the 95th and 99th percentile values, respectively, for these two latency metrics.

**Utilization:** Utilization is defined as the percentage of model instances actively serving an inference request (i.e., *busy*), divided by the total number of instances provisioned in the cluster. A model replica may be in one of two states at any time: *idle* or *busy*. We collect a time-series by sampling this cluster state with an interval of 2 seconds. The mean value within the benchmark duration is taken for each request load test, and tabulated by each system for comparison.

### 5.3 Method

To identify a suitable SLA value for each coding task, we set up the cluster with 6 replicas pre-provisioned per CodeLLM

and invoked each coding task 5 times at a rate of 1.67 req/s using 'power-of-two-choices' scheduling to establish a baseline. We measured the average TTFT across all runs for code completion and code generation, and the average E2E latency for code summarization and code translation. We then set these values as each task's SLA target to calculate Goodput for the load testing experiments in each RQ.

We used CodeXGLUE [28] which contains dataset for all four coding tasks we selected for this study as below.

- code generation - uses Text-To-Code dataset to generate functions using the prompt
- code summarization - uses Code-To-Text dataset to summarize a given Java function
- code completion - uses Code-Completion dataset to predict the line given a code snippet (Python)
- code translation - uses Code-To-Code dataset to translate functions from one language to other (Java to C#)

In the first three RQs, we compare the scheduling capabilities, and in RQ4, we compare both scheduling and scaling capabilities together. We ran multiple load tests with different numbers of requests following a Poisson distribution arrival pattern. For RQ1 and RQ2, we used a request proportion of 90% for TTFT latency-sensitive tasks (code generation and code completion) and 10% for E2E latency-sensitive tasks (code summarization and code translation), as the volume of TTFT latency-sensitive requests is expected to be higher than E2E latency-sensitive requests during development time. For RQ3 and RQ4, we used two different mixtures of request arrival patterns:

- *Pattern 1* with 40% code completion, 40% code generation, 10% code translation and requests 10% code summarization.
- *Pattern 2* with 25% requests for each coding task. The results for *Pattern 2* are shown in the Appendix section.

For *Goodput*, the best performing system should maintain a high Goodput to provide stable service to customers under all load scenarios. When loads increase, the degree to which Goodput drops determines resilience and effective throughput of the system. For *Latency*, we compared the TTFT and E2E latency distributions of each serving system, the best performing system should have a latency distribution curve with little-to-no area exceeding the SLA threshold (under the P99 or P95 metric). Finally, for *Utilization*, we compared the mean utilization of each serving system. The best system should have high utilization throughout the benchmark duration, indicating there were a greater proportion of LLM replicas busy during the benchmark.

## 5.4 RQ1: Can SLA-aware scheduling improve Goodput of long-output coding tasks?

Both code generation and code translation are more likely to generate long outputs. For instance, in code translation, CodeXGLUE data shows more that than 40% of the generated code has 200-400 tokens and more than 70% of the CodeLLM's output are in the 100-400 tokens range.

*5.4.1 Approach.* We deployed both code generation and code translation on all three systems on the cluster. As both code generation and the code translation have two CodeLLM calls (Figure 2), for the first CodeLLM, we used CodeLlama2-7B [44] and StarCoder2-7B [25] as the second CodeLLM. The number of replicas is fixed and pre-provisioned in the cluster; 6 replicas of CodeLlama and 6 of StarCoder2, occupying all 12 accelerator cards.

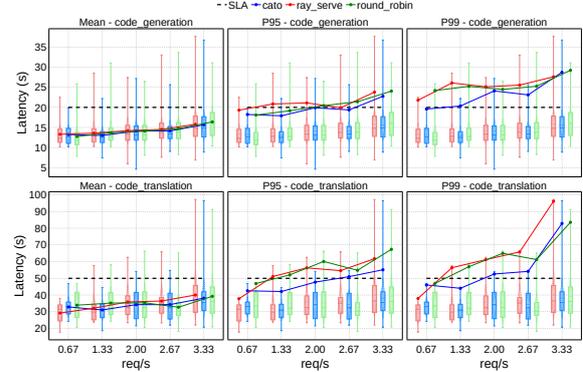

**Figure 4: Comparing mean, P95, and P99 latency measurements for code generation and code translation under load testing.**

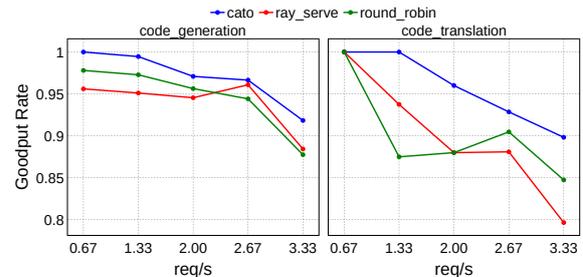

**Figure 5: Goodput measurements for code generation and code translation**

*5.4.2 Results.* Long-output generating tasks such as code translation and code generation benefit from SLA-aware scheduling. For code generation scenario, TTFT P99 was reduced by 2-5 seconds compared with the Ray Serve baseline, at low loads of <2req/s in Figure 4. The P95 latency metric was maintained under the SLA dotted line for CATO, at all but the heaviest load at 3.33 req/s. However, all three systems were not able to maintain P99 under SLA due to the limitation of cluster resources. Similarly, for code translation tasks, we observe that CATO's SLA-aware scheduling provides an improvement in the P95 E2E latency by at least 5-10s compared with Ray Serve and Round Robin.

Figure 5 shows the trend of Goodput rate as the load increases. For code generation, CATO consistently outperforms Ray Serve and Round Robin. Under the heaviest load (3.33 req/s), the Goodput rate of Ray Serve and Round Robin drops from more than 0.95 to below 0.90, while CATO maintains a Goodput rate of 0.9x. This demonstrates that CATO is more stable when handling heavier workloads. For code translation, CATO performs better, maintaining a Goodput rate of about 0.9 under heavy load (3.33 req/s), whereas both Ray Serve and Round Robin see their Goodput rates drop to 0.8 and 0.85, respectively, under heavy load.

In terms of hardware utilization, CATO's SLA-aware scheduler was effective in increasing replica utilization between 3-14% over Ray Serve and Round Robin as seen in Table 1, where once again the heaviest load scenario presents the most observable advantage. From the results presented in Figure 4 and Figure 5, it can be observed that under heavier workloads, the Goodput of code translation declines more abruptly compared to code generation. This phenomenon is likely due to the longer prompts required for code translation, which impose greater demands on computing resources during the prefill stage of model inference. Future research could explore combining CATO's scheduling algorithm with prefill and decoding separations [53], to further improve the performance.

| Requests/s | CATO | Ray Serve | Round Robin |
|---|---|---|---|
| 0.67 | 0.2571 | 0.2360 | 0.2092 |
| 1.33 | 0.3023 | 0.2653 | 0.2378 |
| 2.00 | 0.3250 | 0.2637 | 0.2789 |
| 2.67 | 0.3817 | 0.3187 | 0.3455 |
| 3.33 | 0.4157 | 0.3266 | 0.3293 |

**Table 1: Mean utilization of the cluster when running code generation and code translation**

### 5.5 RQ2: Can SLA-aware scheduling improve Goodput of short-output coding tasks?

Both code summarization and code completion will likely generate short outputs. Typically, code completion generates less than ten tokens and code summarization generates a summary text with 20 to 30 tokens (observed from Code-to-Text dataset from CodeXGLUE).

*5.5.1 Approach.* We deployed both code summarization and code completion on all three systems. We used the same CodeLLM (CodeLlama2-7B) on both coding tasks. Models were pre-provisioned the same as in RQ1.

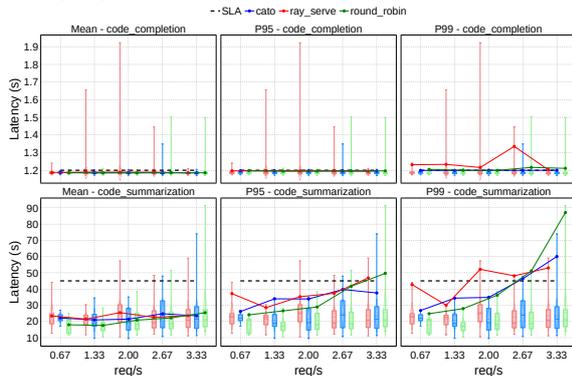

**Figure 6: Comparing mean, P95, and P99 latency measurements for code completion and code summarization under load testing**

*5.5.2 Results.* From results in RQ2, we observe marginal differences in code completion TTFT latency, except the P99 metric was maintained under 1.2 seconds which is the SLA across all load conditions. Ray Serve and Round Robin scheduling policies had significantly more SLA violations and inconsistent latencies in Figure 6. In terms of Goodput, CATO demonstrates an advantage by 1% for code completion and 2-4% for code summarization, against the other 2 scheduling policies in Figure 7.

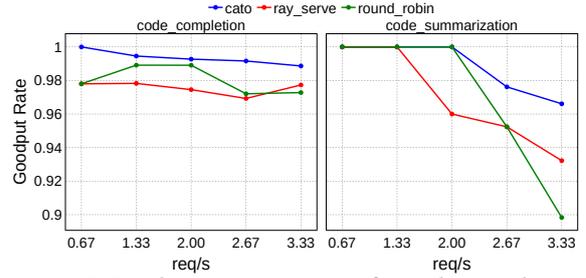

**Figure 7: Goodput measurements for code completion and code summarization**

| Requests/s | CATO | Ray Serve | Round Robin |
|---|---|---|---|
| 0.67 | 0.1599 | 0.1312 | 0.1478 |
| 1.33 | 0.2375 | 0.2717 | 0.2667 |
| 2.00 | 0.3714 | 0.2202 | 0.3056 |
| 2.67 | 0.3776 | 0.3222 | 0.3664 |
| 3.33 | 0.4230 | 0.3322 | 0.3261 |

**Table 2: Mean utilization of the cluster when running code completion and code summarization**

In terms of utilization, we observed no consistent improvement under light to moderate loads (<2.67 req/s), however improvements began when the load was increased to a heavy load of 3.33 req/s, and maintained about a 9% improvement over Ray Serve in Table 2.

### 5.6 RQ3: Can SLA-aware scheduling improve Goodput when serving all types of coding tasks?

In a typical real world scenario, all four types of coding tasks are mixed and will come with different request arrival patterns.

*5.6.1 Approach.* We deployed all four coding tasks (code summarization, code completion, code generation, code translation) on all three systems. Each task has a different SLA (latency requirement) as described in Figure 1. We used the same two CodeLLMs (CodeLlama2-7B and StarCoder2-7B) for long-output tasks used in RQ1 (code generation and code translation) and the same CodeLLM (CodeLlama2-7B) for short-output tasks used in RQ2 (code completion and code summarization). The number of replicas is fixed in the cluster (total of 12 replicas: 6 replicas per CodeLLM model occupying all 12 cards).

| Requests/s | CATO | Ray Serve | Round Robin |
|---|---|---|---|
| 0.83 | 0.3120 | 0.2534 | 0.2261 |
| 1.67 | 0.3586 | 0.2988 | 0.2944 |
| 2.50 | 0.3702 | 0.3199 | 0.3091 |
| 3.33 | 0.4748 | 0.3360 | 0.3596 |

**Table 3: Mean utilization of the cluster when running all four coding tasks**

*5.6.2 Results.* Most SLA violations occurred for code completion at heavier loads Figure 8. For code generation at moderate loads (1.67 req/s), the P95 TTFT has most noticeably improved from 21s to 18.5s (11% improvement). For Goodput, CATO never drops below a Goodput of 0.86 for this task under the heaviest load (3.33 req/s), while Ray Serve and Round Robin had a Goodput of 0.78 (10% relative improvement) and 0.7 (16% relative improvement) in Figure 9 . This can be explained as other policies are not aware of the slack time of enqueued requests, thus they are unable to re-prioritize within a queue and select the best replica for inference, leading to suboptimal workflow latency and overall more SLA violations. In term of utilization, CATO excels by achieving a mean utilization of 0.48, when the Ray Serve had a utilization of 0.34; representing a 41.2% improvement in Table 3.

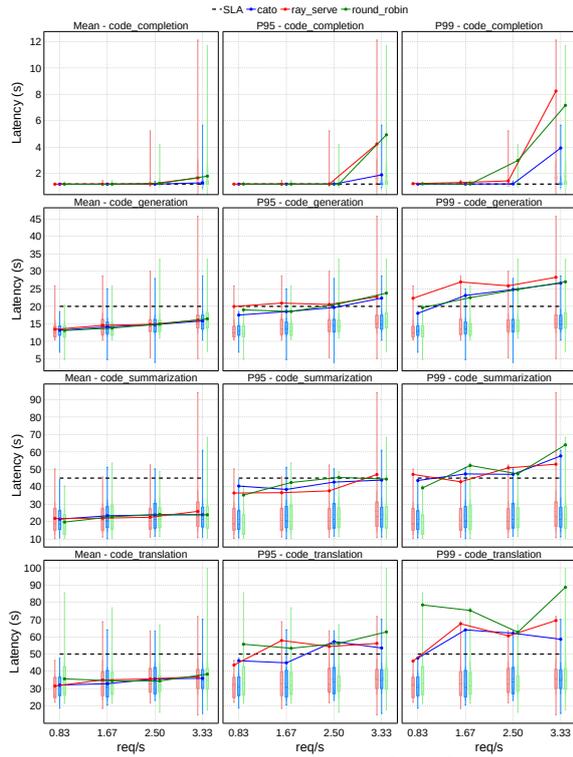

**Figure 8: Comparing mean, P95, and P99 latency measurements for all four coding tasks**

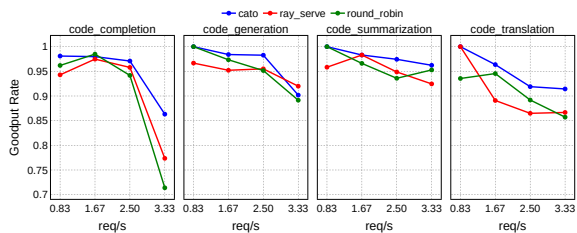

**Figure 9: Goodput measurements for all four coding tasks**

## 5.7 RQ4: Can SLA-aware autoscaling provide a balance of Goodput and Utilization?

The motivation for this research question is to ensure scaling and scheduling policies do not conflict and are compatible. Popular distributed systems such as Ray Serve and Kubernetes [22] monitor live metrics of the cluster to determine when to provision additional replicas, and metrics may include request rate, queue length, hardware utilization[22].

*5.7.1 Approach.* In this experiment, we repeat the RQ3 load test but with autoscaling enabled. The SLA-aware scheduling was paired with SLA-aware scaling (Algorithm 2), and Ray Serve/ Round Robin was paired with maximum queue length based scaling [42]. Unlike the previous experiments, where a fixed number of replicas were already provisioned, the load test was conducted beginning with 1 replica per model, and allowed to scale up over time according to the incoming request load. More specifically, the SLA-aware scaling was set up using the following limits:
- Min number of replicas : 1 replica per CodeLLM (initial state)
- Max number of replicas : 6 replicas per CodeLLM
- SLA violation counts : 1 SLA violation before scaling is triggered (for SLA-Aware autoscaling)
- SLA violation threshold : 1 SLA violation before scaling is triggered (for SLA-Aware autoscaling)

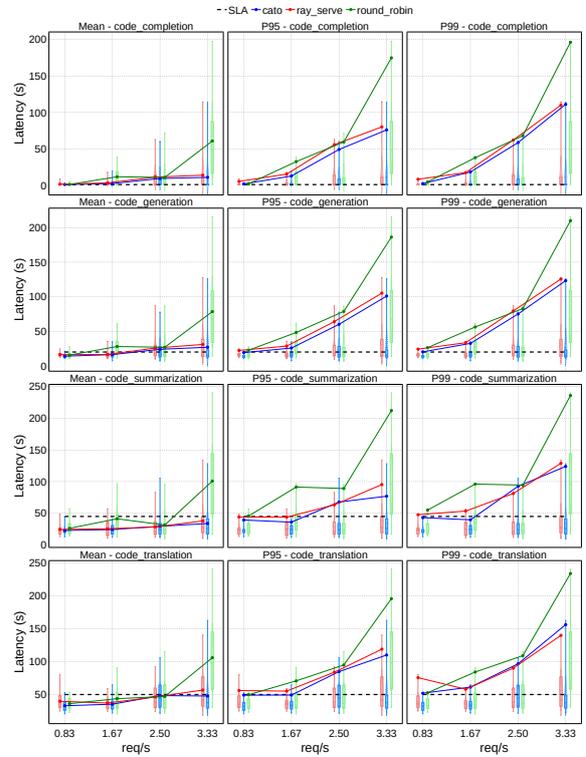

**Figure 10: Comparing mean, P95, and P99 latency measurements for all four coding tasks with both scaling and scheduling**

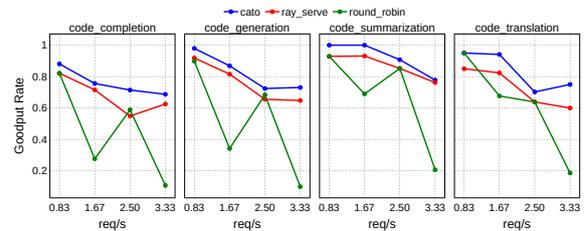

**Figure 11: Goodput measurements for all four coding tasks with both scaling and scheduling**

*5.7.2 Results.* In Figure 10, latencies grew linearly with request arrival rate. This is due to the cold start delay of each replica taking time to provision; the system responsiveness may be adjusted using the exceeded_times and threshold parameter in Algorithm 2 (5). This replica provisioning incurs the expensive model-loading process, which involves data movement of the model weights from NVMe storage into accelerator memory (i.e., a cold start). Despite this, the P95 latencies of the SLA-aware scaling in all code tasks were consistently lower than the alternative Ray Serve and Round Robin policies. In particular, code summarization P95 E2E latency at the moderate load (1.67 req/s) was reduced from 44s to 36s over Ray Serve while still meeting SLA, representing a 18% improvement. And code generation P95 TTFT latencies at the lightest load (0.83 req/s) were reduced from 22s to 19s, improving over Ray Serve by 14%. For the other TTFT-critical workload, code generation, SLA-aware scaling maintains a relatively high Goodput of 0.73 at the heaviest load scenario (3.33 req/s), proving effective and beating Ray Serve at 0.65 (8% improvement) and Round Robin at 0.1 (63% improvement) in Figure 11.

| Requests/s | CATO | Ray Serve | Round Robin |
|---|---|---|---|
| 0.83 | 0.2768 | 0.3185 | 0.3287 |
| 1.67 | 0.4308 | 0.4676 | 0.5182 |
| 2.50 | 0.4351 | 0.5209 | 0.4611 |
| 3.33 | 0.4276 | 0.4580 | 0.5007 |

Table 4: Mean utilization of the cluster when running all four coding tasks with both scaling and scheduling

The mean utilization of CATO in Table 4 is comparatively lower than Ray Serve and Round Robin. This can be explained as replicas are provisioned early before request at the queues of each replica has a chance to grow, as soon as any node executions have fallen behind the allotted slack in the coding task. This behavior of SLA-aware autoscaling can lead to some intervals of idle replicas in the initial scaling phase, but in return, the Goodput across all tasks is consistently better since these replicas are already loaded when a subsequent requests arrive. In practice, this interval can be fine-tuned to further balance based on requirements and stringency of meeting SLA.

## 6 THREATS TO VALIDITY

**External:** In large scale production systems, clusters comprise of over 100 machines. Due to resource constraints, the experiments were only performed on 2 machines, also due to time constraints, each system was only subject to 100-200 requests in a 60-second window, while other trends may emerge at 1k-10k requests. As an external threat, we attempt to mitigate it by separating the analysis, to study the behavior of each system with a) the cluster at steady state in RQ1-3 (replicas pre-provisioned) and b) initial response in RQ4 (autoscale from 1 replica), to provide a scale-agnostic performance analysis.

**Internal:** As an internal threat, we chose to perform the ablation study with scaling and scheduling components of CATO perform actions independently to avoid confounding factors. We further employ random sampling during the metrics collection phase to remove run-to-run variance as much as possible, as the hardware may exhibit noise (such as PCIe bus contention with other processes on the cluster).

## 7 RELATED WORK

There are a limited number of works which address both scheduling and scaling together to meet SLA requirements of coding tasks; scaling is often delegated to external systems like Kubernetes[22] without fine-grained understanding of these performance requirements. Ray Serve [43] is the only other system which provides scaling and scheduling policies to serve LLMs at the *workflow* level, and coding tasks can be described as workflows. Due to this, we chose Ray Serve as the baseline for the experiments.

There are some recent works which address the scheduling aspect in serving tasks. Teola [48] uses topology-aware batching, which is a heuristic approach to optimize the execution of queries, by intelligently grouping requests based on the structure and dependencies of the underlying data flow graph. However, this work has not considered the scaling of model replicas but rather uses a pre-configured number of replicas. ParrotServe [26] has proposed manually annotating applications with different Quality of Service (QoS) expectations, such as latency vs throughput, and the underlying engine processes them according to predefined hard-coded values. However, this work proposes changes to the underlying LLM serving infrastructure (requiring changes to vLLM). The hard-coded QoS is also inflexible as the application will execute either the latency-oriented (output will be limited with the predefined low number of tokens to generate) or throughput-oriented (higher limit on output tokens) configuration, which will not work if the user requires stringent and fine-grained SLA management of each workflow.

There is another field of work which optimizes serving LLM requests with Model-as-as-Service infrastructures (Llumnix [46], Mélange [19]), but since they do not consider entire LLM workflows and additionally do not have autoscaling capability, we cannot directly compare them in our experiments. Frameworks that serve LLM workflows (such as SageMaker[4], Triton [34]) require standalone tools for static resource configuration at deployment time. Once again, developers often use container based orchestrations, such as Kubernetes, with default scaling policies for horizontal scaling[22], which could lead to poor SLA adherence (low Goodput rate) as shown in our experiments.

## 8 CONCLUSION

We propose the Coding Assistant Task Orchestrator (CATO) with SLA-aware scheduling and scaling algorithms for serving CodeLLMs, which coding assistants rely on. We analyze and categorize four types of coding tasks according to their characteristics and respective latency requirements (TTFT vs. E2E latency critical). We evaluate CATO by load testing it against state-of-the-practice systems, such as Ray Serve and Round Robin schedulers. We evaluate each configuration's performance on TTFT and E2E Latency metrics, Utilization, and Goodput. In RQ1, the load test consists of short-output tasks. In RQ2, the load test consists of long-output tasks. In RQ3, all four types of coding tasks were served simultaneously, and the advantages of SLA-aware scheduling carried over across all task types in both TTFT and E2E latency, most noticeably under heavy loads (3.33 req/s). In RQ4, SLA-aware scaling was enabled and compared with request queue-length-based scaling. CATO's SLA-aware configuration improved overall Goodput rate and resource utilization by up to 10% and 41.1%, respectively. In terms of latency, CATO reduced P95 E2E latency by 18% for code summarization tasks, and P95 TTFT for code generation tasks were reduced by 14 %. The experimental results support that Goodput across all mixtures of coding assistant tasks improved when using SLA-aware scheduling and auto-scaling policies. CATO is developed to be generic (not limited to only serving coding tasks), which opens up the possibility of serving other workflows and AI applications. We are planning to explore this in future research.

## 9 DISCLAIMER

Any opinions, findings, conclusions, or recommendations expressed in this material are those of the author(s) and do not reflect the views of Huawei. Also, ChatGPT-4.0 was used for copy-editing. All experiments, analysis, writing, and results were performed by the authors, who also thoroughly reviewed the final content. This complies with IEEE and ACM policies on AI use in publications.

SLA-Awareness for AI-assisted coding                                    Conference'17, July 2017, Washington, DC, USA

# A APPENDIX

Figures 12 to 16 show the measures when we repeated the RQ3 and RQ4 experiments with the *Pattern 2* (where all four coding task requests arrived in equal proportions). Figures 12 to 14 are for RQ3 and Figures 15 and 16 are for RQ4 experiments. From the figures, CATO SLA-aware system still outperforms the others in terms of Latency and Goodput.

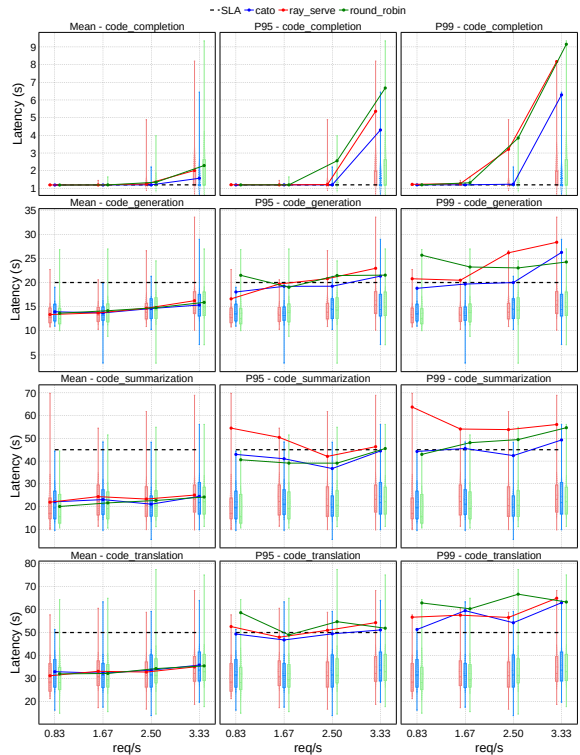

**Figure 12: Comparing mean, P95, and P99 latency measurements for all four coding tasks (Request Pattern 2)**

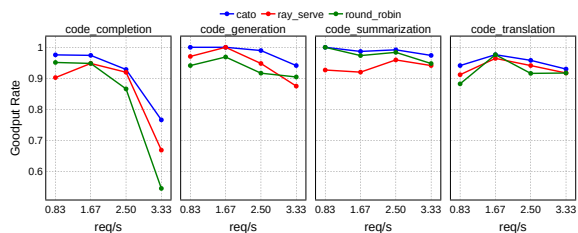

**Figure 13: Goodput measurements for all four coding tasks (Request Pattern 2)**

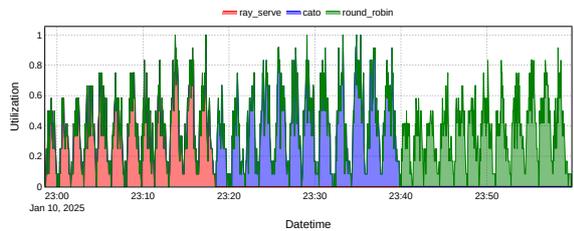

**Figure 14: Overall utilization of the cluster when running all four coding tasks (Request Pattern 2)**

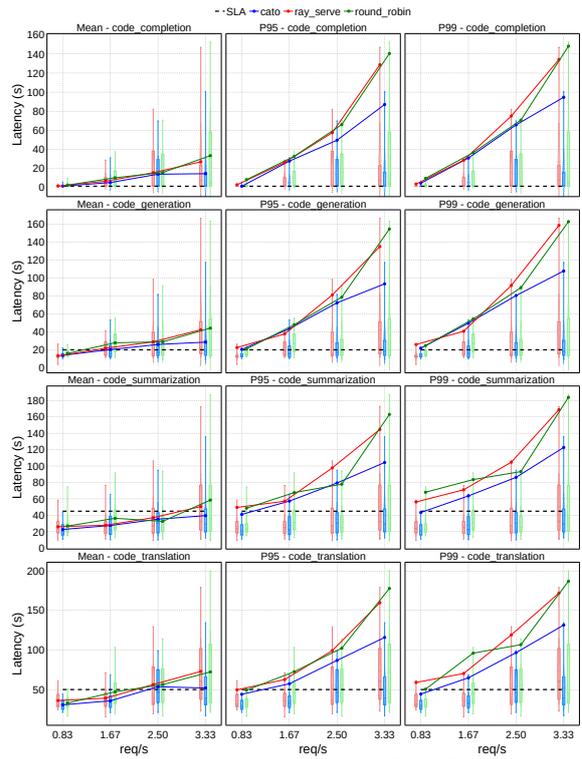

**Figure 15: Comparing mean, P95, and P99 latency measurements for all four coding tasks with both scaling and scheduling (Request Pattern 2)**

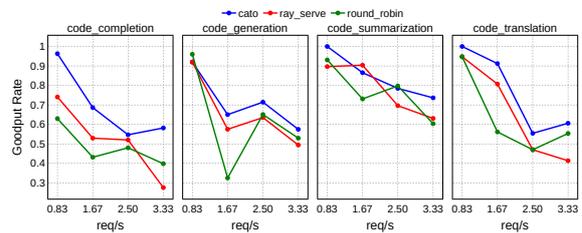

**Figure 16: Goodput measurements for all four coding tasks with both scaling and scheduling (Request Pattern 2)**